\documentstyle[seceq,mbf,wrapft]{ptptex}

\preprintnumber[3cm]{
KUNS-1325\\PTPTeX ver.0.8\\ August, 1997}

\title{
Uniqueness and Non-Uniqueness of Static Vacuum Black Holes in Higher Dimensions
}

\author{
Gary W {\sc Gibbons}, Daisuke {\sc Ida}$^{*}$  
and Tetsuya {\sc Shiromizu}$^{*,**}$
}

\inst{
DAMTP, Centre for Mathematical Sciences, The University
of Cambridge,\\
Wilberforce Road, Cambridge CB3 0WA, United Kingdom 
\\
$^*$Department of Physics, Tokyo Institute of Technology,
Tokyo 152-8551, Japan \\
$^{**}$Advanced Research Institute for Science and Engineering,
Waseda University, Tokyo 169-8555, Japan 
}

\abst{
We prove the uniqueness theorem for asymptotically flat static vacuum black
hole
solutions in higher-dimensional space-times. We also construct
infinitely many non-asymptotically flat regular static black holes
on the same space-time manifold with the same spherical topology.
}

\begin{document}

\maketitle


\makeatletter
\if 0\@prtstyle
\def\asp{.3em} \def\bsp{.26em}
\else
\def\asp{.3em} \def\bsp{.3em}
\fi \makeatother

\section{Introduction}

With the development of string theory, black holes in higher-dimensional
space-times have come to play a fundamental role in physics \cite{BH}.
Furthermore,
the possibility of black hole production
in high energy experiments has  recently been suggested  in the context of
the
so-called TeV gravity \cite{LHC}.
A TeV-size black hole in TeV gravity is small enough to be well approximated
by
an asymptotically flat black hole in higher dimensions.
To predict  phenomenological
results, we need reliable  knowledge about higher-dimensional black
holes.
However, some essential  features of black hole
theory have not so far been fully explored.
Among these, the equilibrium problem for  black holes is
one of most important issues.
The final equilibrium state of the black hole is known to
drastically simplify  in the case of four space-time dimensions,
because of the uniqueness properties  of
static or stationary black hole solutions.
The uniqueness theorem for  the vacuum black hole is well established
in four-dimensional space-times. \cite{4d1,4d2,4d3} (See also
Ref.~\cite{Review} for
comprehensive review)
Although this no hair property is  fundamental to the nature of black holes,
it is at the same time a quite non-trivial result derived from the Einstein
equations.
So far, there is no evidence that in higher dimensions
the final state of the black hole is unique.
Remarkably, five-dimensional stationary vacuum black holes are  not unique;
there is a Myers-Perry solution \cite{BH}, which is a
generalization of the Kerr solution
to arbitrary dimensions, while Emparan and Reall \cite{Reall} have recently
found  five-dimensional rotating black ring solutions with the same
angular momenta and mass but with an  event horizon
homeomorphic to $S^2\times S^1$.
In the static case, such a counter-example has not
yet been presented.
The only known asymptotically flat static vacuum black hole
is the $n$-dimensional hyperspherically symmetric
Schwarzschild-Tangherlini solution \cite{Tangherlini}.
We shall show in what follows that there are no 
others\footnote{After submitting this manuscript to gr-qc, 
we were informed by M. Anderson that a similar proof has been given by Hwang\cite{Hwang}.}. 
Our proof can be extended to charged dilatonic cases\cite{Tess}. 
However, it is interesting to note that, as we shall expand upon below,
if one drops the condition of asymptotic flatness but still
insists that the space-time have the same topology as  the
Schwarschild-Tanghelerlini solution, then the uniqueness property
fails badly: There exist discrete infinities  of solutions.

In four dimensions, there are essentially two ways of proving the
uniqueness  of the Schwarzschild solution.
The first is  Israel's original proof\cite{4d1}, which cannot be
generalized to higher dimensions in a simple manner.
For example, it uses the Gauss-Bonnet theorem to evaluate the
surface integral of the Ricci curvature of each level set of the
length  of the time-like Killing vector field. This  can be
done only in the case that the level set is two dimensional.
Another proof was
given by Bunting and Masood-ul-Alam\cite{4d2}.
This method uses a  corollary of the positive mass theorem
to establish that the
domain of outer communication is spatially conformally flat.
Then an  identity involving  the Bach-Weyl tensor of the spatial geometry
is used to prove that the conformal flatness of the spatial geometry
implies  spherical symmetry.
In higher dimensions, this argument does not hold, since the Bach-Weyl tensor
is no longer  a conformal tensor.

\section{Proof of uniqueness}

The first half of our reasoning is parallel to  Bunting-Masood-ul-Alam's
method, and therefore we shall only  briefly describe this part here.
In general, the metric of an  $n$-dimensional
static space-time has the form
%
\begin{eqnarray}
ds^2=-V^2dt^2+g_{ij}dx^idx^j,
\end{eqnarray}
%
where $V$ and $g_{ij}$ are independent of $t$
and  are regarded as quantities on the $t=[{\rm constant}]$
hypersurface $\Sigma$.
The event horizon $H$ is a Killing horizon
located at the level set $V=0$,
which is assumed to be non-degenerate. In fact non-degeneracy  follows from
Smarr's formula relating the mass, surface gravity and area of the horizon.
We impose the vacuum Einstein equations,
%
\begin{eqnarray}
D_iD^iV=0 \label{einstein1}
\end{eqnarray}
%
and
%
\begin{eqnarray}
{}^{(n-1)}R_{ij}-\frac{1}{V}D_i D_j V=0, \label{einstein2}
\end{eqnarray}
%
where $D_i$ and ${}^{(n-1)}R_{ij}$ denote a covariant derivative
and the Ricci tensor defined on
$(\Sigma, g_{ij})$, respectively.

In asymptotically
flat space-times, one can find an appropriate coordinate system in which
the metric has an asymptotic expansion of the form
\begin{eqnarray}
V = 1-\frac{C}{r^{n-3}}+O(1/r^{n-2})
\end{eqnarray}
and
\begin{eqnarray}
g_{ij}=\left(
1+\frac{2}{n-3}\frac{C}{r^{n-3}}\right)\delta_{ij}+O(1/r^{n-2}),
\end{eqnarray}
where $C=[{\rm constant}]$ represents the ADM mass (up to a constant factor)
and $r:=\sqrt{\sum_i(x^i)^2}$.

Consider the  two conformal transformations
\begin{eqnarray}
\tilde g_{ij}^{\pm}=\Omega_\pm^2 g_{ij},
\end{eqnarray}
where
\begin{eqnarray}
\Omega_{\pm}=\left( \frac{1 \pm V}{2}\right)^{2/(n-3)}.
\end{eqnarray}
Then we have two manifolds ($ \Sigma^{\pm}, g_{ij}^{\pm}$).
On $\Sigma^{+}$, the asymptotic behavior of the metric
becomes
\begin{eqnarray}
\tilde g_{ij}^{+} =\delta_{ij}+ O\left(1/r^{n-2} \right).\label{asymptotic}
\end{eqnarray}
On  $ \Sigma^{-}$, we have
\begin{eqnarray}
\tilde g_{ij}^{-} dx^i dx^j & = &
\frac{(C/2)^{4/(n-3)}}{r^4}\left(dr^2+r^2d\Omega^2_{n-2}\right)+O(1/r^5)\nonumber\\
& = & (C/2)^{4/(n-3)}\left(dR^2+R^2d\Omega^2_{n-2}\right)+O(R^5),\nonumber\\
\end{eqnarray}
where $d\Omega^2_{n-2}$ denotes the round sphere metric and $R:=1/r$ has
been defined.
Pasting $( \Sigma^{\pm}, g_{ij}^{\pm})$ across the
level set $V=0$ and adding a point $\{p\}$ at $R=0$,
we can construct a complete regular surface
$\tilde\Sigma= \Sigma^{+} \cup \Sigma^{-}\cup \{p\}$.
It can be shown that the Ricci curvature on $\tilde\Sigma$ vanishes.
Furthermore, Eq.~(\ref{asymptotic}) implies that
the total mass also vanishes on $\tilde\Sigma$.
As a consequence of the positive mass theorem \cite{PET,Marika},
such a surface $\tilde\Sigma$ must be flat.
In other words, $g_{ij}$ can be written in the conformally flat form
\begin{eqnarray}
g_{ij}&=&v^{4/(n-3)}\delta_{ij},\\
v:&=&\frac{2}{1 + V}.
\end{eqnarray}
Then, the Einstein equation (\ref{einstein1})  reduces  to
the Laplace equation on the  Euclidean $(n-1)$-space
\begin{eqnarray}
\nabla_0^2 v=0, \label{laplace}
\end{eqnarray}
where $\nabla_0$ denotes the flat connection.

On the other hand, since the function $V$ is a harmonic function,
it can be used as a  local  coordinate in a
neighbourhood $U \subset \Sigma$
of each connected component of the horizon.
Let $\{x^A\}$ be coordinates on level sets of $V$
such that their trajectories are orthogonal to each level set.
Then, the metric on $\Sigma$ can be  written in the form
\begin{eqnarray}
g_{ij}dx^i dx^j=\rho^2 dV^2+h_{AB}dx^A dx^B.
\end{eqnarray}
From Eq.~(\ref{einstein2})
and the conformal flatness of $\Sigma$,
the Riemann tensor on $\Sigma$ becomes
\begin{eqnarray}
{}^{(n-1)}R_{ijkl}=\frac{2}{n-3}\frac{1}{V} \left( g_{i[k}D_{l]}D_j V
-g_{j[k}D_{l]}D_iV \right).
\end{eqnarray}
The $n$-dimensional Riemann invariant can be calculated using the 
formulae
\begin{eqnarray}
{}^{(n)}R_{ijkl}&=&{}^{(n-1)}R_{ijkl},\\
{}^{(n)}R_{0i0j}&=&VD_iD_jV,\\
D_i D_j V&=& \frac{1}{\rho}k_{ij}-\frac{2}{\rho^2} n_{(i} D_{j)} \rho
+\frac{\partial_V \rho}{\rho^3} n_i n_j,
\end{eqnarray}
where $n=\rho dV$ and $k_{AB}$ denote the unit normal and the extrinsic
curvature
of the $V=[{\rm constant}]$ surface, respectively.
Thus, we obtain
\begin{eqnarray}
{}^{(n)}R_{\mu\nu\lambda\rho}{}^{(n)}R^{\mu\nu\lambda\rho}& = &
{}^{(n)}R_{ijkl}{}^{(n)}R^{ijkl}+
4{}^{(n)}R^{0i0j}{}^{(n)}R_{0i0j} \nonumber \\
& = &
\frac{4(n-2)}{(n-3)V^2}( D_i D_j V)( D^i D^j V) \nonumber \\
&  = & \frac{4(n-2)}{(n-3)V^2\rho^2}\left[k_{AB}k^{AB}+k^2 +2 {\cal D}_A \rho
{\cal D}^A \rho\right]
\label{kretchmann}
\end{eqnarray}
where ${\cal D}_A$ denotes the covariant derivative on each level set of
$V$.

The requirement that the event horizon $H$ be a regular surface
leads to the condition
\begin{eqnarray}
&&k_{AB}|_H=0,\\
&&{\cal D}_A\rho|_H=0.
\end{eqnarray}
In particular, $H$ is a totally geodesic surface in $\Sigma$.

Let us consider how the event horizon
is embedded into the base space $(\tilde\Sigma,\delta_{ij})$.
We can adopt the following expression for the base space metric:
\begin{eqnarray}
\delta_{ij}dx^idx^j=\tilde \rho^2 dv^2+\tilde h_{AB}dx^A dx^B.
\end{eqnarray}
The event horizon is located at the $v=2$ surface, $\tilde H$.
Then, the extrinsic curvature $\tilde k_{AB}$ of the level set $v={\rm
constant}$
can be expressed as
\begin{eqnarray}
\tilde k_{AB}= v^{-2/(n-3)}k_{AB}+\frac{2}{n-3}
\frac{v^{(n-1)/(n-3)}}{\rho} \tilde h_{AB}.
\end{eqnarray}
Thus we obtain
\begin{eqnarray}
\tilde k_{AB}= \frac{2^{(2n-4)/(n-3)}}{(n-3)\rho|_H} \tilde h_{AB}.
\end{eqnarray}
on $\tilde H$. In other words, the embedding of $\tilde H$ into the
Euclidean $(n-1)$-space
is totally umbilical.
It is known that such an embedding must be hyperspherical; \cite{sphere}
that is,
each connected component of $\tilde H$ is a geometric sphere with a certain
radius $r_0$ determined by
the value of $\rho|_H$.
The embedding of a hypersphere into the Euclidean space is known to be rigid
\cite{rigid},
which means that we can always locate one connected component of
 $\tilde H$ at the $r=r_0$ surface of $\tilde \Sigma$
without loss of generality.

Now we have a boundary value problem for  the
 Laplace equation on the base space
$\Omega:=E^{n-1}\setminus B^{n-1}$ with the Dirichlet boundary conditions.
The system is characterized by a parameter $\rho|_H$, which fixes the radius of
the inner boundary $\tilde H=\partial B^{n-1}$.
The field equation is the Laplace equation (\ref{laplace}) with the
Dirichlet condition
\begin{equation}
v|_{\tilde H}=2
\end{equation}
and the asymptotic decay condition
\begin{equation}
v=1+O(1/r^{n-3}),
\end{equation}
with $r\rightarrow +\infty$.
Let $v_1$ and $v_2$ be solutions of this boundary value problem.
Integration of the Green identity
\begin{eqnarray}
\nabla_0^2\left[\frac{1}{2}(v_1-v_2)^2\right]&=&\nabla_0 (v_1-v_2)\cdot
\nabla_0 (v_1-v_2)+(v_1-v_2)\nabla_0^2 (v_1-v_2)
\end{eqnarray}
over $\Omega$ gives
\begin{eqnarray}
&&\left(\oint_{r=+\infty}-
\oint_{\tilde H}\right)(v_1-v_2)\frac{\partial}{\partial r}
(v_1-v_2)dS\nonumber\\
&=&\int_\Omega |\nabla_0 (v_1-v_2)|^2dV.
\end{eqnarray}
Since the surface integral vanishes due to the boundary condition imposed,
the integrand
in the volume integral must be identically zero: $v_1-v_2=[{\rm constant}]$.
Then, from the boundary condition, two solutions must be identical
($v_1=v_2$),
which shows the uniqueness of the solution of this boundary value problem
if the horizon is connected:
{\em The only $n$-dimensional asymptotically flat static vacuum
 black hole with non-degenerate regular event horizon
is the Schwarzschild-Tangherlini family.}

In fact, we assumed that the horizon is connected to obtain the
above theorem. However, we can remove this assumption as follows.
Let us consider the evolution of the level surface in
Euclidean space. From the Gauss equation in Euclidean space, we
obtain  the evolution equation for the shear $\tilde \sigma_{AB}
:=\tilde k_{AB}-\tilde kh_{AB}/(n-2)$:
\begin{eqnarray}
\mbox \pounds_{\tilde n}  \tilde \sigma_{AB}
& = & \tilde \sigma_{A}^{~C} \tilde \sigma_{CB}
+\frac{1}{n-2}\tilde h_{AB} \tilde \sigma_{CD} \tilde \sigma^{CD} 
-\frac{1}{\tilde \rho}\biggl( {\tilde {\cal D}}_A {\tilde {\cal D}}_B
-\frac{1}{n-2}\tilde h_{AB} {\tilde {\cal D}}^2  \biggr)\tilde \rho.
\end{eqnarray}
Using $\Delta_0 v=0$, we can derive the equation for
${\tilde {\cal D}}_A {\rm ln}\tilde \rho$ as
\begin{eqnarray}
\mbox \pounds_{\tilde n} {\tilde {\cal D}}_A {\rm ln}\tilde \rho=
\tilde k {\tilde {\cal D}}_A {\rm ln}\tilde \rho +{\tilde {\cal D}}_A \tilde
k.
\end{eqnarray}
For the trace part of $\tilde k_{AB}$, we have
\begin{eqnarray}
\mbox \pounds_{\tilde n} \tilde k = -\tilde
\sigma_{AB}^2-\frac{1}{n-2}k^2-\frac{1}{\tilde \rho}
{\tilde {\cal D}}^2 \tilde \rho
\end{eqnarray}
and
\begin{eqnarray}
\mbox \pounds_{\tilde n} {\cal D}_A \tilde k ={\tilde {\cal D}}_A \mbox
\pounds_{\tilde n}
\tilde k + ({\tilde {\cal D}}_A {\rm ln}\tilde \rho)(  \mbox \pounds_{\tilde
n}
\tilde k).
\end{eqnarray}
From the above equations, we can see that
\begin{eqnarray}
\tilde \sigma_{AB}=0, ~~{\tilde {\cal D}}_A \tilde \rho=0~~~{\rm
and}~~~{\tilde {\cal D}}_A
\tilde k =0;
\end{eqnarray}
that is, each level surface of $v$ is totally
umbilic, and hence spherically symmetric, which implies
that the metric is isometric to the Schwarzshild-Tangherlini solution.
This is, of course, a local result, since we consider only the region
containing no saddle points of the harmonic function $v$.
To obtain the global result, we need a further assumption, such as
analyticity.
However, the assumption that there is no saddle point may be  justified as
follows.
At a saddle point $\rho=0$, the level surface of $v$ is multi-sheeted;
that is,
the embedding of the level surfaces is singular there.
One can find at least one level surface such that $k_{AB}\ne 0$ near the
saddle point.
Then, Eq.~(\ref{kretchmann}) implies that the saddle point is singular.
If the horizon is not connected,
this naked singularity must  exist to
 compensate for the gravitational attraction
between black holes.

\section{Non-uniqueness}

We have shown that if space-time is assumed to be asymptotically
flat then the only regular static black hole  is given
by the Schwarzschild-Tangherlini metric. However other,
non-asymptotically flat solutions may be obtained by replacing
the metric of the round $(n-2)$-sphere by any other Einstein manifold
whose Ricci-curvature has the same magnitude as that of a unit round
$(n-2)$-sphere. In particular we can replace the round metric on
$S^5,S^6,\dots, S^9$ by the infinite sequences of Einstein metrics
found recently by Bohm \cite{Bohm}. These have the form
\begin{eqnarray}
d\theta^2 + a^2(\theta ) g_p + b^2(\theta) g_q,
\end{eqnarray}
where $g_m$ is  the unit round
metric on the $m$-sphere, $p+q=n-3$ and neither $p$ nor $q$ is 1.
In general, these metrics are inhomogeneous with
isometry groups $SO(p+1)\times SO(q+1)$.
The round metric on $S^{n-2}$ is given by $a=\sin \theta$ and $b=\cos
\theta$. In addition, Boehm demonstrated the existence of infinite
sequences
of smooth Einstein  metrics that converge to  singular metrics of
finite
volume. By a theorem of Bishop \cite{Bishop}, the volume
of these metrics is always less than that of the round metric.
It follows that for fixed temperature, the associated static black
holes  always have smaller Bekenstein-Hawking entropy than the
Schwarzschild-Tangherlini black hole.
For this reason, we believe  that these metrics are all unstable.
Presumably, they decay to  the Schwarzschild-Tangherlini solution.

Clearly, Bohm metrics may also be used in D3-branes solutions
and applied to the AdS/CFT correspondence.
The volume may then be related to
the central charge of a conformal field theory \cite{Gubser}.
In this way we obtain
a further connection between  the  entropy of horizons, geometry  and
the central charges of
quantum field theories, where now Bishop's theorem provides a universal
bound for the central charge.

\section*{Acknowledgements}

TS's work is partially supported by the Yamada Science Foundation and 
Grants-in-Aid for Scientific Research from the Ministry of Education, Science, 
Sports and Culture of Japan(Nos. 13135208, 14740155 and 14102004).

\end{document}